\documentclass[twocolumn,aps,epsf,showpacs,amssymb,amsmath,prb]{revtex4}
\usepackage{graphicx}
\usepackage{epsfig}

\newcommand{\la}{\langle}
\newcommand{\ra}{\rangle}

\newcommand{\nn}{\nonumber}

\newcommand{\be}{\begin{eqnarray}}
\newcommand{\ee}{\end{eqnarray}}

\begin{document}
\preprint{APS/123-QED}
\title{Superfluid Insulator Transitions of Hard-Core Bosons on the Checkerboard Lattice}
\author{Arnab Sen$^1$, Kedar Damle$^1$, and T Senthil$^{2,3}$}

\affiliation{%
$^1$Department of Theoretical Physics,
Tata Institute of Fundamental Research,
1 Homi Bhabha Road, Mumbai 400005, India\\
$^2$ Department of Physics, Indian Institute of Science, Bangalore 560012, India\\
$^3$ Department of Physics, Massachussetts Institute of Technology, Cambridge, MA 02139,
USA
}%

\date{\today}

\begin{abstract}
We study hard-core bosons on the checkerboard lattice with nearest neighbour unfrustrated hopping $t$ and `tetrahedral' plaquette charging energy $U$. Analytical arguments and Quantum Monte Carlo simulations lead us to the conclusion that the system undergoes a zero temperature ($T$) quantum phase transition from a superfluid phase at small $U/t$ to a large $U/t$ Mott insulator phase with $\rho$ = 1/4 for a range of values of the chemical potential $\mu$. Further, the quarter-filled
insulator breaks lattice translation symmetry in a characteristic four-fold ordering pattern, and occupies a lobe of finite extent in the $\mu$-$U/t$ phase diagram. A Quantum Monte-Carlo study slightly away from the tip of the lobe provides evidence for a direct weakly first-order superfluid-insulator transition away from the tip of the lobe. While analytical  arguments leads us to conclude that the transition {\em at} the tip of the lobe belongs to a different landau-forbidden second-order universality class, an extrapolation of our numerical results suggests that the size of the first-order
jump does not go to zero even at the tip of the lobe.
\end{abstract}

\pacs{75.10.Jm 05.30.Jp 71.27.+a}
\vskip2pc

\maketitle

\section{Introduction}

Experiments on a variety of condensed matter systems such as under-doped high-temperature superconductors, strongly correlated quasi-two dimensional organic compounds like $\kappa$-(ET)$_2$Cu$_2$(CN)$_3$\cite{Kanoda} and the triangular lattice antiferromagnet CsCuCl$_4$ \cite{Coldea} all point to the possibility of realizing genuinely new phases of matter outside of standard paradigms such as fermi liquid theory of normal metals, spin-wave theory of magnetically ordered states, and the BCS theory of superconductivity. There has been a great deal of activity in the recent past aimed at developing theoretically consistent descriptions of such genuinely new phases of matter---examples include various proposals for topologically ordered liquid states of strongly-correlated systems, and their gauge-theoretic description \cite{Read_Sachdev,Wen,Senthil0,Motrunich_Senthil, Balents_Fisher_Girvin,Motrunich0, Hermele_Balents_Fisher}. 

A generic feature of these `exotic' phases is that their low lying excitations are described in terms of quasiparticles that carry unusual quantum numbers and are best thought of as fractions of the usual electron or magnon excitations. These fractionalized quasiparticles interact with each other by emergent gauge forces, and have an independent existence only in such exotic phases where the gauge forces scale to zero at large distances effectively `deconfining' the fractionalized quasiparticles.

In closely related work, Senthil~{\it et. al.}~\cite{Senthil_etal} have also developed a theory of a class of second order quantum phase transitions that fall outside of the rubric of Landau's theory of phase transitions. In the usual Landau theory, the transition is described by analyzing fluctuations of the order parameter field that encodes the sharp distinction between the two phases. In certain situations where the phases on either side of the transition break different symmetries,  Senthil {\it et. al.} demonstrated that this usual description fails, in that it would erroneously rule out the possibility of a direct second order transition in the generic case.

Such a direct transition is actually possible, and its properties are best understood in terms of new fractionalized degrees of freedom (that can be thought of as fractions of the quasiparticle excitations of  the adjoining phase) that are liberated at precisely the critical point~\cite{Levin_Senthil}.
This has sparked interest~\cite{Lee, Melko0,Melko1,Melko2,Sandviklatest} in theoretical studies of relatively simple model systems where one can check for the presence of such unusual phases of matter and  non-landau `deconfined' quantum phase transitions with numerical studies.

In recent work~\cite{Damle_Senthil}, two of the present authors argued that a $S=1$ antiferromagnet with isotropic exchange $J$ and strong uniaxial single-ion anisotropy $D$ on the kagome lattice would support an unusual spin-nematic state in zero field, and a spin-density wave magnetization plateau state for a range of non-zero magnetic fields along the easy-axis. Both phases and the transition between them can be modeled by a Hamiltonian for hard-core bosons with unfrustrated hopping $t$ and frustrated nearest-neighbour repulsive interactions $U$\cite{Damle_Senthil}; the
nematic state corresponds to a ordinary superfluid state of bosons at half-filling, while the magnetization plateau state corresponds to a $1/3$ filled charge-density wave insulator  that occupies a lobe in the $\mu$-$U/t$ phase diagram (where $\mu$, the chemical potential for bosons, is linearly related to the applied magnetic field, and $U/t \sim D^2/J^2$).

As the two phases break different symmetries, the transition between them has the possibility of being an unusual non-Landau transition~\cite{Isakov_Kim0}. Numerical studies of the transition close to the tip of the lobe~\cite{Damle_Senthil,Isakov_Kim1} reached the tentative conclusion
that the transition between the two phases may indeed be a direct second-order transition with unusual exponents, although more recent work  on larger sizes seems to see  faint signatures of a very weak first-order transition close to the tip~\cite{Isakov_Kim1}.

\section{The Model and schematic phase diagram}

Motivated in part by this intriguing transition seen on the Kagome lattice, and its possible relevance to the physics of anisotropic antiferromagnets,  we consider here a closely analogous model of hard-core bosons on the checkerboard (planar pyrochlore) lattice, with the aim of shedding further light on the physics of such systems (As our work was nearing completion, other work on similar {\it fermionic} models~\cite{Fulde1,Fulde2,Shtengel} and somewhat different SU(2) symmetric
spin models~\cite{Nussinov1,Nussinov2} has also appeared in the literature).

\begin{figure}
\includegraphics[width=\hsize]{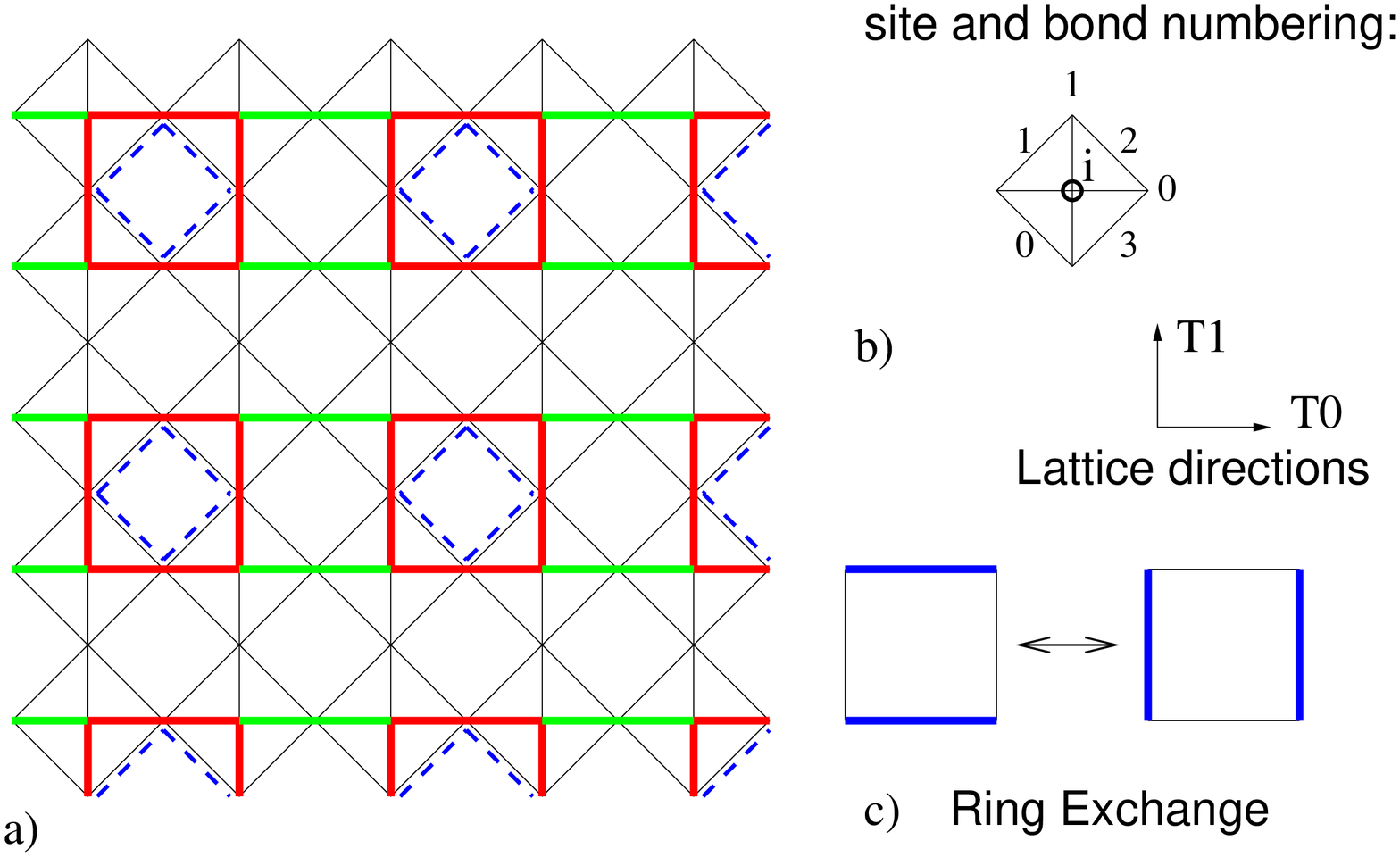}
      \caption{(Color online). a) Periodic Checkerboard lattice and the square lattice whose bonds pass through the checkerboard sites. The basic corner-sharing units of this lattice are `tetrahedral' plaquettes centered at the sites of the square net (shown in b)). The presence of a boson on a site of the checkerboard lattice can be represented as a `dimer'
 covering the corresponding link of the underlying square lattice. In the plaquette ordered state, red squares resonate via the ring-exchange process (shown in c)) while the other links are empty ($n_{i}$ = 0). The dotted bonds indicate higher kinetic energy in the plaquette state. In the alternate columnar state, dimers cover all green edges but not red ones.}
      \label{lattice}
  \end{figure}

The checkerboard or planar pyrochlore lattice is made up of four site plaquettes which can be thought of as projections of a tetrahedron onto a plane, and are centered at the sites of an underlying square net (Fig~\ref{lattice}). The bosonic particles live on the vertices of this `tetrahedron' (which lie on the centers of the links of the square net). The Hamiltonian for our boson hubbard model reads

\be
\label{microham}
H = -t\sum_ {\la ij\ra}(b^{\dagger}_{i}b_{j} + h.c.)~~~~~~~~~~~~~~~~~~~~~~~~~\nn \\
\mbox{~~~~~~~~~~~~~}+ U\sum_{p}(n_{1p}+n_{2p}+n_{3p}+n_{4p}-f)^2
\ee

where $\la ij\ra$ refers to the nearest neighbour links of the checkerboard lattice, $b^{\dagger}_i$ ($b_{i}$) creates (destroys) a particle on site $i$, $t > 0$ denotes the nearest neighbour hopping, and $U > 0$ denotes the repulsion energy between bosons on the same `tetrahedral' plaquette $p$.
Note that $n_{1p}, n_{2p}, n_{3p}$ and $n_{4p}$ are the boson number operators of the sites of the $p^{th}$ such tetrahedral plaquette (which can equally well be indexed by the site of the underlying square lattice around which it is centered), and the
parameter $f$ serves as a reduced chemical potential that tunes the particle density in the ground state.

Before we begin a detailed discussion of our numerical results, it is useful to delineate the gross features of the phase diagram using fairly general and reliable arguments. To this end, we begin in the limit of vanishingly small repulsive interaction $U \ll t$.
As $U/t\rightarrow 0$, the hopping term in the Hamiltonian dominates and its unfrustrated nature immediately leads us to the conclusion that the ground state is a featureless superfluid. The opposite limit when $U/t\rightarrow\infty $ and the potential energy term dominates is best understood by first going over into the `atomic' limit of isolated tetrahedra ($t=0$) and then analyzing the properties of the perturbation expansion in $t/U$. In the atomic limit, the density on each tetrahedron is controlled by the reduced chemical potential $f$:

For $0 < f < 1/2$, the ground state has no particles, while for $1/2 < f < 3/2$, the preferred density is one particle per tetrahedron
(which translates to a density of $\rho=1/4$), and for $3/2 < f < 5/2$, each tetrahedral plaquette is half-filled. Further, the physics for $f >2 $ can be mapped exactly to the physics at $4-f$ by a canonical transformation (specific to {\em hard-core} bosons) which is best written in
terms of the equivalent pseudospin-1/2 operators $S^x-iS^y \equiv b$, $S^z \equiv n-1/2$: $S^y \rightarrow -S^y$, $S^z \rightarrow -S^z$, $S^x \rightarrow S^x$. Thus the density as a function of $f$ consists of a series of plateaus with
a preferred number of particles per tetrahedral plaquette, and in view of the above, it suffices to consider the physics of the $\rho=1/4$
and $\rho=1/2$ plateaux to obtain a complete account of the physics.

A crucial feature of these plateaus (except the trivial cases with no particles and a fully filled lattice) is that they represent properties of a manifold of degenerate ground states with macroscopically large degeneracy. For concreteness, we focus here on the plateau with one particle per tetrahedron (our results for the $\rho=1/2$ plateau will be discussed separately).  A convenient representation of the ground state manifold is to associate a dimer on a link of the underlying square lattice  with the presence of a particle on the corresponding site on the checkerboard lattice. Each state in the ground state manifold then corresponds to a perfect dimer cover of the underlying square lattice.

The $t$ term introduces quantum tunneling between these dimer states, which lifts the classical degeneracy. In the dimer language, the leading order effective Hamiltonian in the $U/t\rightarrow\infty $ limit is a quantum dimer model with a ring-exchange term (Fig~\ref{lattice}) of $O(t^2/U)$ that flips two parallel horizontal (vertical) dimers on any {\it flippable} plaquette to vertical (horizontal) dimers, and acts entirely within the classical ground state manifold. 
\begin{figure}
\includegraphics[width=0.9\hsize]{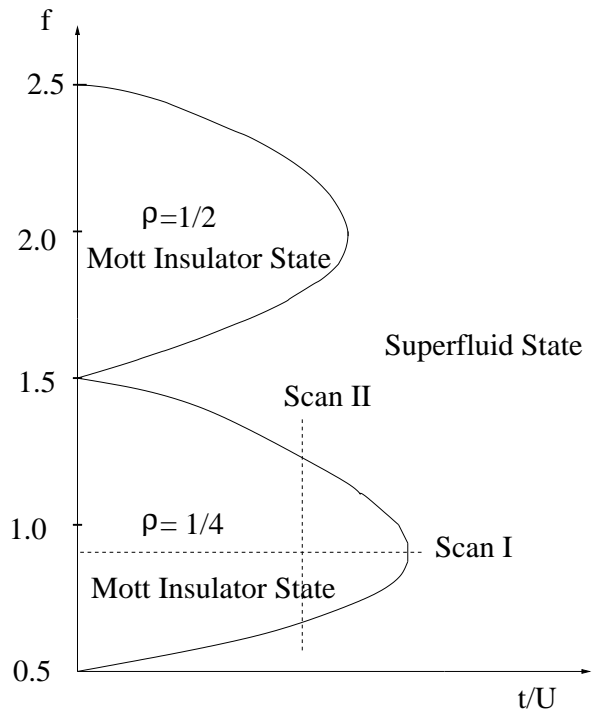}
      \caption{Schematic phase diagram for the transition from superfluid state to Mott insulator state at $\rho$ = 1/4. Scans I and II are discussed in text.}
      \label{phasediagram}
  \end{figure}

Quantum dimer models with ring-exchange on individual plaquettes of two dimensional bipartite lattices (here a square lattice) quite generally have crystalline ground states in which the spatial arrangement of dimers break lattice symmetry~\cite{Sachdev_Vojta}. In the square lattice case, one expects either a plaquette or columnar type Valence Bond Solid (VBS), in which there is crystalline order at $\vec{Q}_1 = (\pi,0)$ and $\vec{Q}_2 = (0,\pi)$ ~\cite{Syljuasen} (lattice directions defined in Fig~\ref{lattice}). Thus, at large but finite $U/t$, one has a Mott insulating lobe in the $f-U/t$ phase diagram with density pinned to $\rho$ = 1/4 for a finite range of values around $f \sim 1$.

As has been discussed earlier by Fisher~{\it et. al.} in their seminal analysis~\cite{Fisher_Fisher} of superfluid insulator transitions in bosonic lattice models, the contour of fixed density $\rho = 1/4$ is expected to terminate on the tip of the lobe, while all contours with density different from this value skirt around the Mott insulating lobe. Consequently,  the tip of the lobe of the $\rho$ = 1/4 Mott insulator represents a special point along the locus of superfluid - Mott insulator transitions---indeed, the low-energy theory at the tip is expected to have an additional particle-hole symmetry arising from identical excitation energies for quasi-particle and quasi-hole excitations.

This sort of reasoning leads to the schematic phase diagram shown in Fig~\ref{phasediagram}, but is silent on the nature of the quantum phase transition lines between the Mott insulating lobes and the superfluid state.  Since one of the phases is expected to break spatial lattice translation symmetry, and the other superfluid state breaks the internal $U(1)$ symmetry associated with particle number conservation, a Landau-type description of the competition between these two very different order parameters would generically predict (along the $\rho=1/4$ contour) either a first order phase transition, or an intermediate regime with both phases or an intermediate regime with neither order. 

An intermediate phase with neither order would correspond to a liquid phase of bosons in two dimensions at zero temperature. As no consistent scenario for such a phase is known in two dimensions
in the present context, we reach the tentative conclusion that such a phase is unlikely. An intermediate phase with both orders would correspond to a lattice supersolid, and several examples of this are known in the literature~\cite{Supersolid1,Supersolid2,Supersolid3,Supersolid4,Supersolid5,Supersolid6}.  A direct second order phase transition between the two broken symmetry phases would require ``fine-tuning'' to a multicritical point in a Landau type description. However, as has been shown recently~\cite{Senthil_etal} by Senthil {\it et al.}, conventional Landau theory can fail in certain closely related bosonic models in which quantum mechanical Berry phase effects produce a direct second order phase transition.
\begin{figure}
\includegraphics[width=\hsize]{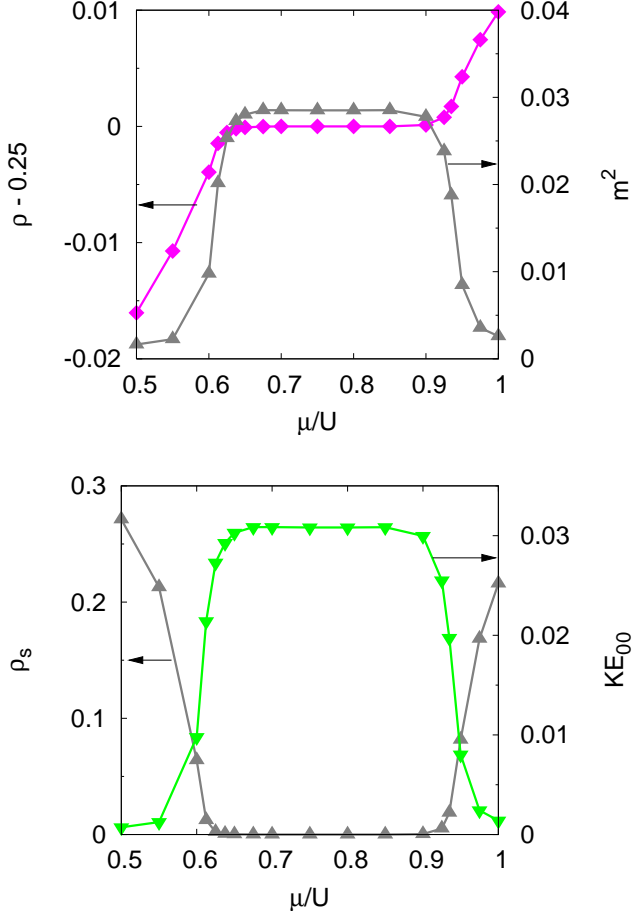}
      \caption{(Color online). Vertical scan (II) showing mott insulator state at $U/t = 3.50$ and $T = t/10$ for size $L =24$. Here $m^2$ = $|\psi_{c}|^2$, with $\psi_c$ defined as in Eqn~({\protect{\ref{def_of_orderparameters}}}). Note that the magnitude of the
      bragg peak in the kinetic energy correlator is also non-zero in the plateau state; this is exemplified by the
      data shown here (${\mathrm{KE}}_{00}$) on the bragg peak in the correlator of kinetic energy along  bond-type labeled $0$
      in Fig~{\protect{\ref{lattice}}}. Error bars are smaller than the symbol sizes in all the cases.}
      \label{plateau}
      \end{figure}

 \section{Numerical study}
  Our goal here is to study this model more closely using the well-documented stochastic series expansion (SSE) Quantum Monte Carlo (QMC) method~\cite{SSEprb,SSEpre,SSEmath} to access the phase diagram (Fig~\ref{phasediagram}), and then compare our numerical results with the expectations built on the previous work of Senthil {\it et. al.}. For this purpose, we have performed a detailed study along two scans in the $U/t$-$f$ phase diagram: Scan I varies $U/t$ at a fixed value of $f$---this value has been chosen after several preliminary exploratory scans to ensure that  it intersects the locus of phase transitions {\it{close}} to the tip of the lobe. The second scan (II) was performed at fixed $U/t$ such that one encounters an insulating phase for an appreciable range of $f$, and can probe the nature of the ordered state in somewhat more detail at a point relatively deep in the insulator. In addition, to probe
 nature of the particle-hole symmetric transition in more detail, we have
  also performed a detailed study of (crystalline) order-parameter statistics for a sequence of critical points approaching
  the tip of the lobe. 
\begin{figure}
\includegraphics[width=0.8\hsize]{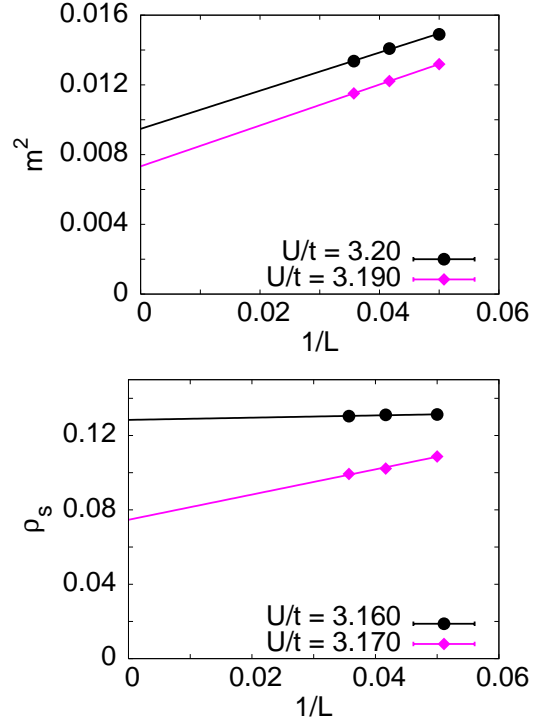}
\caption{(Color online). Finite size extrapolations establishing the presence of superfluid and charge density wave ordered phases. Here $f = 0.875$ and $T = t/15$. Error bars are smaller than the symbol sizes.}
\label{extrapolations}
\end{figure}

Most of our data is on $L \times L$ samples (where $L \times L$ is the number of unit cells) where $L$ ranges from 20 to 48. We use standard SSE estimators to calculate the superfluid density $\rho_{s}$, the equal time correlator of the density $S_{\alpha \beta}(\bf{q})$ = $\langle n_{\alpha\tau}({\bf{q}}) n_{\beta\tau}(-{\bf{q}}) \rangle$ and the static structure factor of the density at zero frequency $\chi_{\alpha\beta}(\bf{q})$ =  $\langle \int{d\tau n_{\alpha\tau}({\bf{q}}) n_{\beta0}}({-\bf{q}}) \rangle$ where $n_{\alpha\tau}({\bf{q}})$ = $(1/L^2) \sum_{i} n_{\alpha \tau}({\bf{r}}_i)exp(i{\bf{q}}{\bf{r}}_i)$, $n_{\alpha \tau}({\bf{r}}_i)$ being the boson number operator at imaginary time $\tau$, and all site types ($\alpha$ = 0,1)  are assigned the coordinates of the  corresponding site (i) of the square net (Fig~\ref{lattice}). We also calculate the static correlators for the `kinetic energy' $K_{l}$ = $(b^{\dagger}_i b_j + h.c.)_l$ on the link $l$, $S^{\alpha \beta}_K ({\bf{q}})$ = $\langle \int d\tau K_{\alpha \tau}({\bf{q}}) K^{\dagger}_{\beta 0}({\bf{q}}) \rangle$, where $K_{\alpha \tau}({\bf{q}})$ = $(1/L^2) \sum_{i} K_{\alpha \tau}({\bf{r}}_i) exp(i {\bf{q}} {\bf{r}}_i)$, $K_{\alpha \tau}({\bf{r}})$ being the kinetic energy operator at imaginary time $\tau$ (all bond types ($\alpha$ = 0,1,2,3) are assigned the coordinates of the  corresponding site $(i)$ of the square net as in Fig~\ref{lattice}).

To verify the presence of the Mott insulator lobe in the phase diagram, we performed vertical scans at constant $U/t$ for different values of $f$. One particular scan (Scan II) at $U/t$ = 3.5 is shown here (Fig~\ref{plateau}). The scan clearly shows the existence of the mott insulator over an appreciable region in the phase diagram. 
As is clear from Fig~\ref{extrapolations}, the insulating phase has Bragg peaks in the structure factor $S_{00}({\bf{q}})$  [$S_{11}({\bf{q}})$] and the static susceptibily $\chi_{00}({\bf{q}})$ [$\chi_{11}({\bf{q}})$] at wavevectors $(\pi,0)$ [$(0,\pi)$] which tend to a non-zero value in the the $L \rightarrow \infty$ limit.  In contrast, there are no appreciable Bragg peaks in the cross-correlator or off-diagonal static susceptibility $S_{01}({\bf{q}})$  [$\chi_{01}({\bf{q}})$].  The wavevectors of these peaks point to the presence of either plaquette and columnar order in the insulator.

To obtain a better characterization of this ordering, we measure two complex order parameters sensitive to the spatial symmetry breaking
\be
\psi_c &=& n_{0 \tau}(\vec{Q}_1) + i n_{1 \tau}(\vec{Q}_2) \nonumber \\
\psi_k &=& (K_{2\tau}(\vec{Q}_1) + K_{3 \tau}(\vec{Q}_1) - K_{0
 \tau}(\vec{Q}_1) - K_{1 \tau}(\vec{Q}_1)) \nonumber \\ &+&  i(K_{1 \tau}(\vec{Q}_2) + K_{2 \tau}(\vec{Q}_2) - K_{0 \tau}(\vec{Q}_2) - K_{3 \tau}(\vec{Q}_2)) \nonumber \\
 \label{def_of_orderparameters}
\ee
where $\vec{Q}_1 = (\pi, 0)$ and $\vec{Q}_2 = (0 , \pi)$.

This spatial symmetry breaking can now be analyzed using a Landau theory written in terms of the order parameters $\psi_c$ and $\psi_k$.  For our purposes here, it suffices to focus on just the ``potential energy'' part of the landau theory, and leave out all terms involving spatial
gradients. Terms in the landau theory are constrained by the requirement of invariance under the symmetries of the system, and the relevant symmetry operations in our case are the discrete operations of the space group of the square lattice, {\it i.e.} $\pi/2$ rotations, translations, reflections and inversion. The action of these on the order parameters is given in Table~\ref{tab:sym}.
\begin{table}[htbp]
  \centering
\vspace{0.5cm}
  \begin{ruledtabular}
  \begin{tabular}{l|c|c|c}
    Operation & Coordinates & $\psi_c$ transformation & $\psi_k$ transformation \\
 \hline
$R_{\pi/2}$ & $x_i \rightarrow \epsilon_{ij} x_j$ & $\psi_c \rightarrow i \psi_c$ & $\psi_k \rightarrow i \psi_k$ \\
$T_y$ & $y \rightarrow y+1$ & $\psi_c \rightarrow \psi_c^{*}$ & $\psi_k \rightarrow \psi_k^{*}$ \\
$T_x$ & $x \rightarrow x+1$ & $\psi_c \rightarrow -\psi_c^{*}$ & $\psi_k \rightarrow -\psi_k^{*}$\\
${\cal R}_x$ & $x \rightarrow -x$ & $\psi_c \rightarrow -\psi_c^{*}$ & $\psi_k \rightarrow -\psi_k^{*}$\\
${\cal R}_y$ & $y \rightarrow -y$ & $\psi_c \rightarrow \psi_c^{*}$ & $\psi_k \rightarrow \psi_k^{*}$\\
${\cal I}$ & $r \rightarrow -r$ & $\psi_c \rightarrow -\psi_c$ & $\psi_k \rightarrow -\psi_k$
\end{tabular}
\end{ruledtabular}
\caption{Transformations of the order parameters $\psi_c$ and $\psi_k$ on the square lattice under the discrete symmetry generators of the lattice. Here $i = 1,2=x,y$ is a spatial index.}
\label{tab:sym}
\end{table}

\begin{figure}
\includegraphics[width=\hsize]{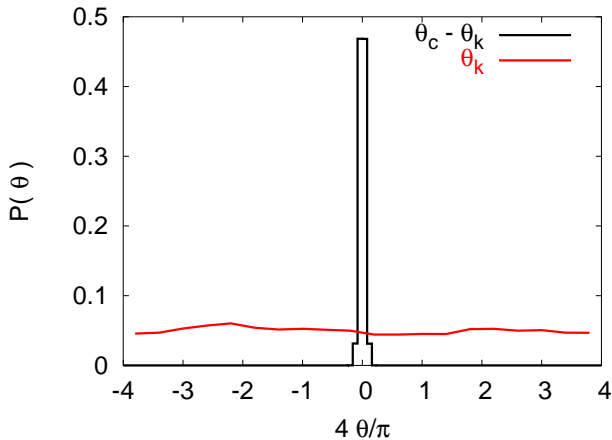}
      \caption{(Color online). Histograms of the relative phase of the two order parameters and the absolute phase of the kinetic energy order parameter characterizing the insulator state at $U/t$ = 3.50, $f$ = 0.875 and $T$ = $t$/15 for size $L$ = 28.}
      \label{landau}
  \end{figure}

Writing all terms consistent with the symmetries of the model (up to fourth order), we obtain the Landau energy functional
\be
{\it{L}}_{pot} & =& {\it{F}}(|\psi_c|^2 , |\psi_k|^2) + C_{\theta_c}(\psi_c^4 + \psi_c^{*4}) + C_{\theta_k}(\psi_k^4 + \psi_k^{*4})\nonumber \\
&+& C_{\theta_{ck}}(\psi_c \psi_k^* + \psi_c^* \psi_k) \; , \nonumber \\
\ee
where $F(x,y)$ is an analytic function of its arguments. 
The nature of ordering (plaquette versus columnar) is determined by the signs of the coefficients $C$ of terms in the landau free energy which fix the relative as well as absolute phases of the two order parameters. Our results on the statistics of the phases of the various order parameters (shown in Fig~\ref{landau} for a point deep in the insulating phase) can be modelled by taking $C_{\theta_{ck}}$ to be negative. However, the signs of $C_{\theta_c}$ and $C_{\theta_k}$ cannot be inferred from these histograms as no statistically significant bias is apparent in the value of the absolute phase.

\begin{figure}
\includegraphics[width=\hsize]{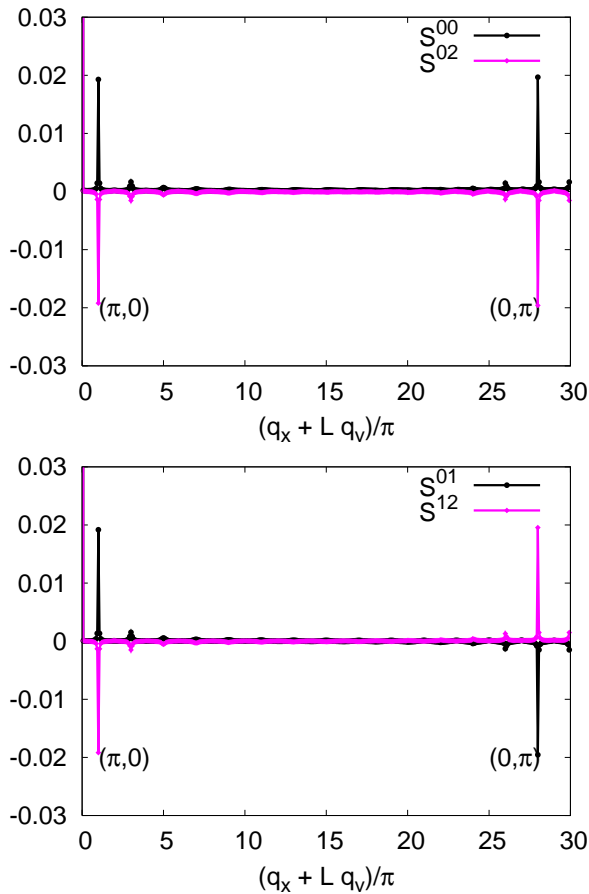}
\caption{(Color online). Plots of $Re(S_{K}^{00}(\vec{q}))$,$Re(S_{K}^{02}(\vec{q}))$, $Re(S_{K}^{01}(\vec{q}))$ and $Re(S_{K}^{12}(\vec{q}))$ for $L = 28$, $U/t = 3.190$, $f=0.875$ and $T = t/15$. The presence of the extrema at wavevectors $(\pi,0)$ and $(0,\pi)$ with the correct signs is consistent with the distribution of K.E. shown in Fig~\ref{lattice}. Error bars are smaller than the symbol sizes.}
\label{kebragg}
 \end{figure}

As the absolute phase is too weakly pinned to usefully discriminate between the plaquette and columnar ordering possibilities, we need to rely on indirect evidence to obtain the nature of the ordering pattern: In this regard, the presence of clearly discernible Bragg peaks in the kinetic energy correlators that survive in the thermodynamic limit is suggestive of plaquette type order (corresponding to a positive sign for the coefficients $C_{\theta_c}$ and $C_{\theta_k}$) since the simplest caricatures of columnar
ordering have essentially no kinetic energy ordering associated with the spatial symmetry breaking.
Indeed, the nature of Bragg peaks seen in the real part of the static correlators of kinetic energy, $S_{K}^{\alpha \beta}(\vec{q})$, is completely consistent with expectations based on the simplest variational wavefunction for
the plaquette insulator. More specifically, the plaquette ordered variational wavefunction denoted pictorially in
Fig~\ref{lattice} leads directly to the following behaviour: $Re(S_{K}^{00}(\vec{q})), Re(S_{K}^{11}(\vec{q})), Re(S_{K}^{22}(\vec{q})), Re(S_{K}^{33}(\vec{q}))$ have maxima at $(\pi , 0)$ and $(0, \pi)$, $Re(S_{K}^{01}(\vec q)), Re(S_{K}^{23}(\vec q))$ have maximum at $(\pi,0)$ and minimum at $(0,\pi)$, $Re(S_{K}^{03}(\vec q)), Re(S_{K}^{12}(\vec q))$ have minimum at $(\pi,0)$ and maximum at $(0,\pi)$ and $Re(S_{K}^{02}(\vec q)), Re(S_{K}^{13}(\vec q))$ have minima at $(\pi,0)$ and $(0,\pi)$ (the other correlators can be obtained from the above ones trivially). Precisely
this behaviour is seen in the QMC data shown in Fig~\ref{kebragg}.

  \begin{figure}
\includegraphics[width=\hsize]{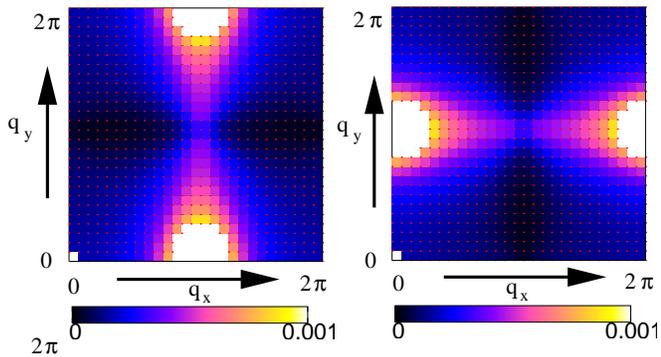}
\caption{(Color online). Static susceptibilities $\chi_{00}({\bf{q}})$ and $\chi_{11}({\bf{q}})$ for $L = 28$, $U/t = 3.180$, $f = 0.875$ and $T = t/15$. In both panels $q_x$ and $q_y$ vary from $0$ to $2\pi$. The white regions are because values of $\chi_{00}$ and $\chi_{11}$ $>$ 0.001 are set to zero.}
\label{dipolar1}
\end{figure}

 In the superfluid phase, there are no Bragg peaks that survive the thermodynamic limit, even in the immediate vicinity of the transition to the insulator. However, equal time correlator  and static structure factor both show an interesting `dipolar' structure in the superfluid phase near the transition (Fig~\ref{dipolar1}). These dipolar correlations do not seem to be associated with the quantum phase transition itself, and persist into the insulating state as well.
 It is therefore instructive to look more closely at the static susceptibilities $\chi_{00}({\bf{q}})$ and $\chi_{11}({\bf{q}})$ and ask whether the dipolar nature is just due the strong local constraint introduced by a large $U$ term.
 
 More specifically, if we are at $f \sim 1$ and temperatures high enough to wash out the coherent quantum dynamics associated with the kinetic energy term, one would expect static density correlators to match (up to prefactors) the predictions for equal time quantities of a classical dimer model on the square lattice (viewing each particle as a dimer on the corresponding link of the underlying square lattice). The dipolar structure seen at lower temperatures can then be viewed as remnants of this essentially classical physics as the system crosses over into a coherent quantum regime.

To check this scenario, we fit the static structure factor to a functional form closely related to the classical result expected for the equal time correlator of classical dimers on the square lattice:
\be
\chi_{00}({\bf{q}}) &=& \frac{a(\cos^2(q_y/2)+\Delta^2)}{\cos^2(q_x/2) + \cos^2(q_y/2)+\Delta^2} \nonumber \\
\chi_{11}({\bf{q}}) &=& \frac{a(\cos^2(q_x/2)+\Delta^2)}{\cos^2(q_x/2) + \cos^2(q_y/2)+\Delta^2}
\label{fiteqn}
\ee
For the critical state of fully-packed dimers, one expects $\Delta^2=0$, and we allow for a non-zero $\Delta^2$ to account for slight density deviations from quarter filling.

\begin{figure}
\includegraphics[width=\hsize]{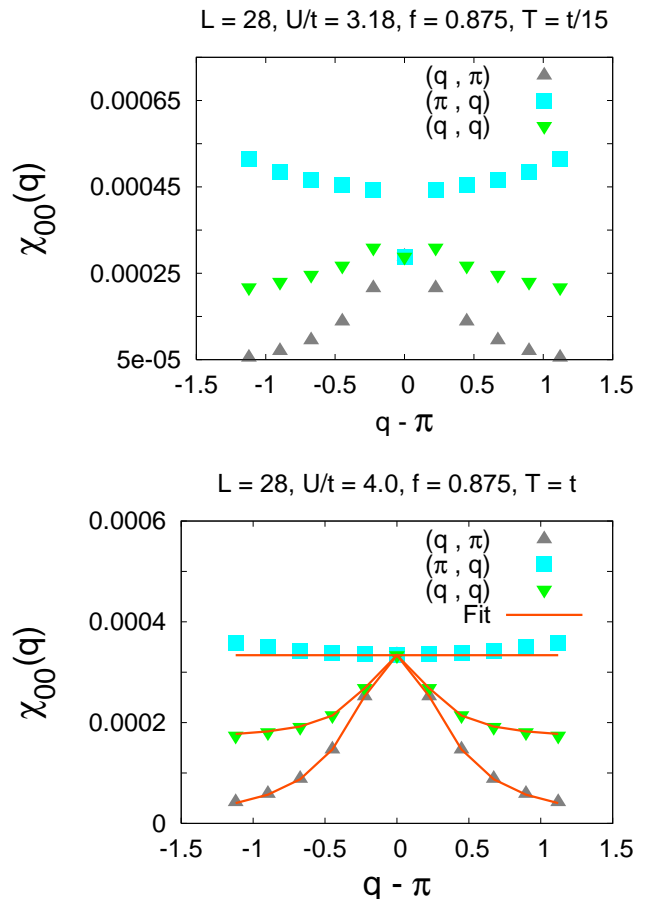}
      \caption{(Color online). Top panel: The static susceptibility $\chi_{00}({\bf{q}})$ plotted along three cuts in momentum space for $L = 28$, $U/t = 3.180$, $f = 0.875$ and $T = t/15$. Bottom panel: The same plotted at $U/t=4$, $f=0.875$ and higher temperature $T = t$, showing an excellent fit to the dipolar form Eqn~({\protect{\ref{fiteqn}}}) with parameters $a= 0.00033$ and $\Delta^2 = 0.0388$. Error bars are smaller than the symbol sizes.}
   \label{fitdipolar}
  \end{figure}

As is clear from Fig~\ref{fitdipolar}, the dipolar structure in the higher temperature data fits remarkably well to our functional form, with a reasonably small value of $\Delta^2 \approx 0.0388$ reflecting a slight deviation of the density from quarter filling. Furthermore, consistent with our expectation above, the low temperature ($\beta t = 15$) data of Fig~\ref{fitdipolar} also displays qualitatively similar features, although the classical form (Eqn~\ref{fiteqn}) does not fit the data as well. 

\begin{figure}
\includegraphics[width=\hsize]{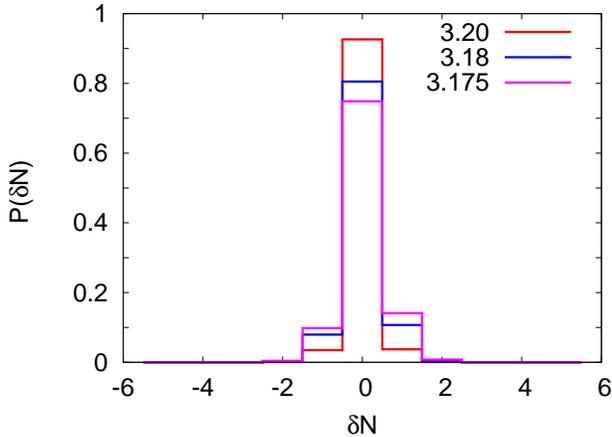}
      \caption{(Color online). Histograms of the density deviation from quarter-filling in the transition region of scan I for size $L=28$. Deviations
      are quantified in terms of the total number of particles by which the system deviates from quarter-filling, and the different
      curves are labeled by the corresponding values of $U/t$ at constant $f=0.875$ and $T=t/15$.}
      \label{phsymmetry}
  \end{figure}

We  now turn our attention to the nature of the quantum phase transition at the tip of the $\rho = 1/4$ lobe.
 We perform Scan I (Fig~\ref{phasediagram}) at a constant aspect ratio of $\beta/L$ = $15/28$. We also compare data at different sizes at a range of inverse temperatures $\beta$ ranging from 5 to 15 to ascertain whether the $T \rightarrow 0$ limit is reached for a given size. The value of $f=0.875$ for this scan was chosen after some trial and error to satisfy two criteria: Firstly, this appears to be close to  the value at which the insulator has maximum extent. Secondly, the system is close to being particle-hole symmetric in the transition region for this value of $f$ (see Fig~\ref{phsymmetry}). 

\begin{figure}
\includegraphics[width=0.9\hsize]{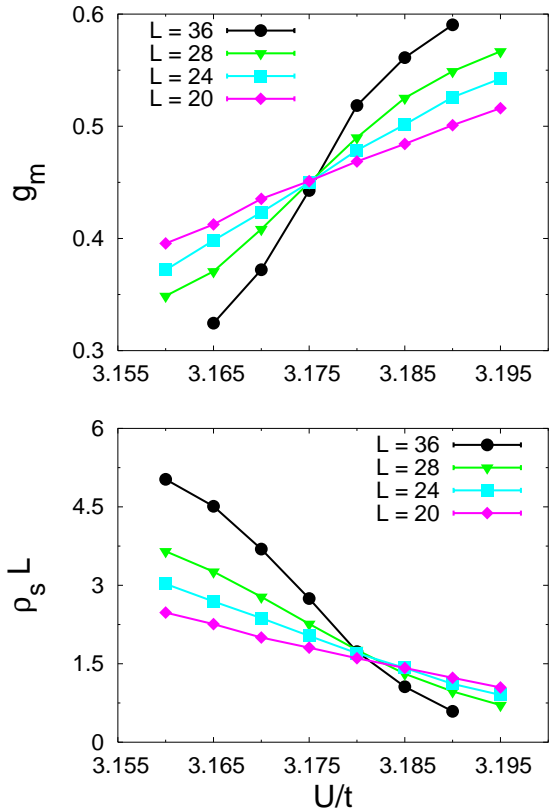}
      \caption{(Color online). Crossing of the Binder cumulant $g_m$ = $1 - \langle m^4 \rangle/(3\langle m^2 \rangle^2)$ where $m^2$ = $|\psi_c|^2$, and $\rho_s$L for different system sizes at $f$ = 0.875 and $\beta/L$ = 15/28. Error bars are smaller than the symbol sizes.}

      \label{scanscaling1}
  \end{figure}

To determine the location of the critical point quantitatively (under the assumption of a scale invariant second-order critical point), we calculate the Binder cumulant $g_{m}$ and $\rho_s L$ for different system sizes at a constant aspect ratio of $\beta/L$ = 15/28. For a continuous phase transition, scaling predicts that these two quantities may be written as $g_m$ = $F_{g_m}(L^{1/ \nu} ((t/U)_c - (t/U)), \beta/L^z)$ and $\rho_s L^z$ = $F_{\rho_s}(L^{1/ \nu} ((t/U)_c - (t/U)), \beta/L^z)$ near the critical point at sufficiently low temperature, where $F_{g_m}$ and $F_{\rho_s}$ are the appropriate scaling functions. Hence if the transition is continuous with $z$ = 1, both the quantities should exhibit the same crossing point in data taken for various $L$ at a fixed aspect ratio $\beta /L$ when plotted versus $U/t$. The reasonably sharp crossings and their locations indicate that
there may be a continuous transition at $U/t \approx 3.178 \pm 0.002$ (to be precise, we obtain a transition at $U/t \approx 3.180$ based on the crossing of $\rho_sL$, while the binder cumulant of the order parameter $m^2$ yields a crossing point at $U/t \approx 3.176$ (see Fig~\ref{scanscaling1})). The crossings shift very little when we consider the constant temperature runs at $T$ = $t$/15 (suggesting that we have reached the asymptotic low temperature regime at these sizes). However, it appears finite-size effects may be more significant, since we do notice a slight tendency of the superfluid and density wave transitions to move closer to each other if we look at the crossings of data at successively larger sizes. Our conclusion based on available data at these
smaller sizes is thus in favour of a single transition near the tip of the lobe, with no appreciable region of phase coexistence.

\begin{figure}
\includegraphics[width=0.9\hsize]{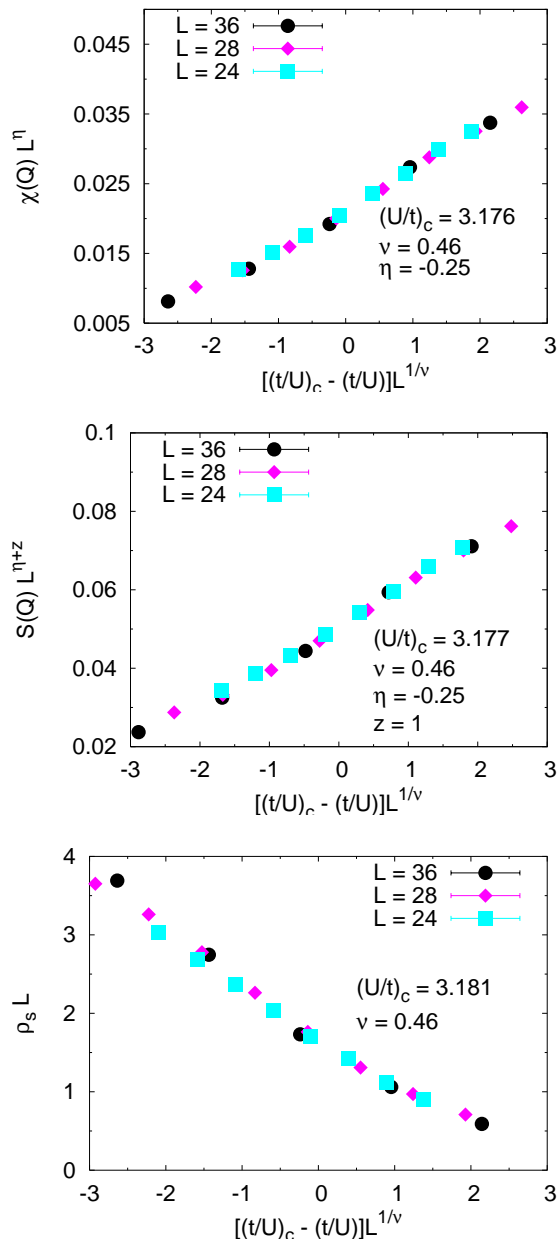}
      \caption{(Color online). Scaling collapse of the static structure factor, equal time correlator and superfluid density at $f = 0.875$ and $\beta/L$ = 15/28. Error bars are smaller than the symbol sizes.}
      \label{collapse2}
  \end{figure}

To further explore the possibility that the transition is actually continuous and not first order, we do a scaling collapse of $\rho_s L$, $\chi(Q)$ and $S(Q)$ (see Fig~\ref{collapse2}), where $\chi(Q)$ = ($\chi_{00}(\vec{Q}_1)$ + $\chi_{11}(\vec{Q}_2)$)/2 and $S(Q)$ = ($S_{00}(\vec{Q}_1)$ + $S_{11}(\vec{Q}_2)$)/2 (the averaging is done to improve statistics). For a direct continuous transition, $\chi(Q)$ = $L^{-\eta} F_{\chi}(L^{1/ \nu}((t/U)_c - (t/U)), \beta/L^z)$ and $S(Q)$ = $L^{-\eta - z} F_{S}(L^{1/ \nu}((t/U)_c - (t/U)), \beta/L^z)$ where the critical exponents are the same as in the $\rho_s L^z$ scaling relation. These scaling relations can also be used to check whether $z$ = 1 by extracting $\eta$ from the scaling collapse of $\chi (Q)$ and then using it for collapsing the $S(Q)$ data. From the scaling collapse of the data, we estimate that $\nu \approx 0.46 \pm 0.05$ and $\eta \approx -0.25 \pm 0.12$ where the error bars are obtained by attempting scaling collapse of the data with different values of the exponents. The scaling collapses are certainly consistent with a direct transition with $z$ = 1, especially since the value of $\nu$ obtained from the superfluid and bragg peak data very nearly coincide within their error bars.

\begin{figure}
\includegraphics[width=\hsize]{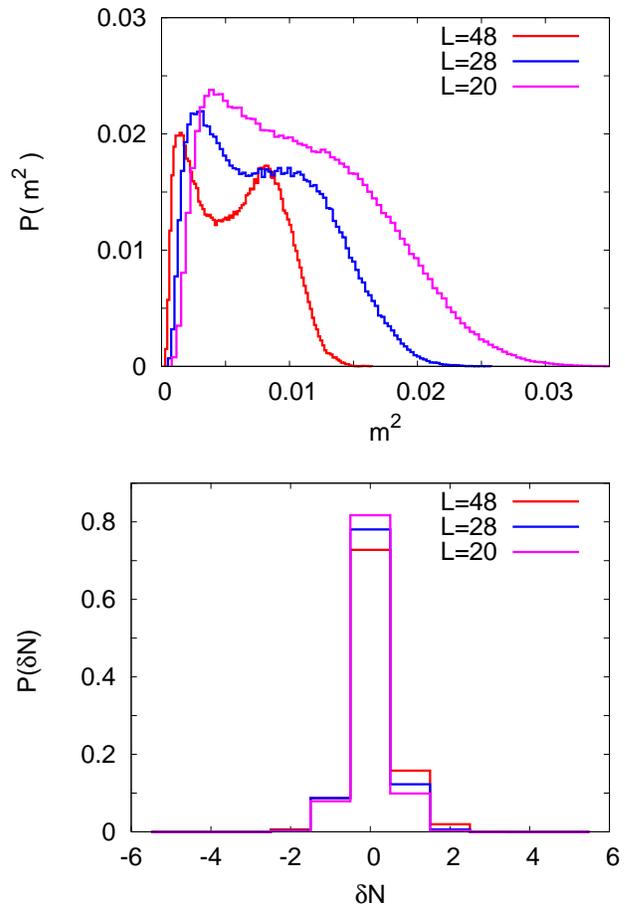}
      \caption{(Color online).  Distribution of the order parameter $m^2$ for different system sizes at $U/t=3.1775$, $f=0.875$ and $\beta/L=15/28$. The double-peaked structure becomes more prominent at larger system sizes, as does the deviation from particle-hole
      symmetry.}
      \label{firstorderproof}
  \end{figure}

While our scaling collapse does seem reasonable, the negative value of $\eta$ obtained from these fits leads  us to speculate that the transition for this particular value of $f=0.875$ may actually be very weakly first order---this is because a negative value of $\eta$ is ruled out on quite general grounds at any conformally  invariant critical point. To confirm this suspicion, we look at the histograms of the charge density wave order parameter $m^2$. They show a characteristic double-peaked structure when one goes to large system sizes indicating that the transition is weakly first order (see Fig~\ref{firstorderproof}).
Thus, based on available evidence, we conclude that the transition close to the tip of the lobe is very weakly first order. The first order nature becomes more apparent at larger sizes $L \ge 48$. It should be noted  that it is precisely at these large sizes that the deviation from particle-hole symmetry at criticality also becomes more evident.
  This correlation in the scaling with size of deviation from particle-hole symmetry and double-peaked nature
  of the order parameter histogram might suggest that the weak first order jump is tied intimately to the slight deviation from particle-hole symmetry.
  
  In order to test this hypothesis, we have located a series of critical points along the phase boundary in the vicinity
  of the tip, and obtained detailed statistics of the charge-density wave order parameter for this sequence of critical
  points. If this hypothesis were valid, one would expect that the double-peak in the order parameter histogram
  at the largest size would become successively less pronounced and extrapolate to zero upon approaching
  the tip of the lobe. Simultaneously, the deviation from quarter filling would also extrapolate to zero.
\begin{figure}
 \includegraphics[width=\hsize]{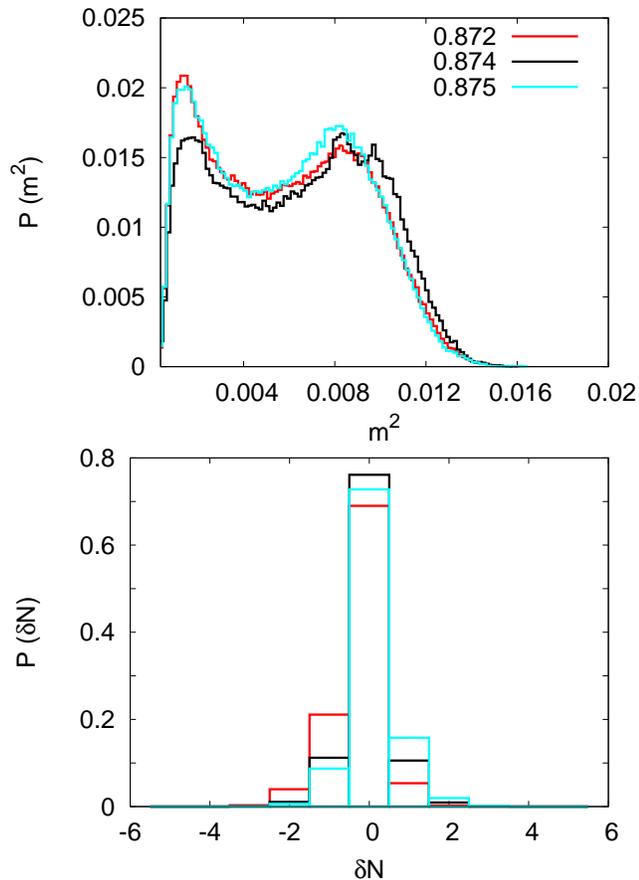}
     \caption{(Color online). Distribution of the order parameter $m^2$ and histogram of deviation of the density
      from quarter filling for $L=48$, $\beta/L=15/28$. The legend indicates the value of $f$ for each of
     these critical points.}
      \label{firstorderattip}
  \end{figure}
  
  However, our numerical results {\em do not} support this hypothesis. This is clear from Fig~\ref{firstorderattip}, where
  we show data at three critical points for our largest system size $L=48$: Of these, one is hole-rich, the other particle rich, and the third is
  perfectly particle-hole symmetric to within our error bars. As is clear from the accompanying histogram
  of the charge-density wave order parameter, there is no significant smearing of the double-peak
  structure upon approaching the particle-hole symmetric transition.
 
 Thus, our numerical results lead us to conclude that although the transition at the tip
 appears to be `close' to a landau forbidden second order transition, it is actually a very weakly first
 order transition as far as we can tell with the largest sizes and lowest temperatures available to us. 
 As we now demonstrate below, an analytical study of the problem leads us to
 the conclusion that the effective low energy theory for the particle-hole symmetric transition
 is precisely the easy-plane NC{\it{CP}}$^1$ model of Ref~\cite{Senthil_etal}. Our conclusion of a weakly
 first-order transition then suggests that more work is needed to fully understand the physics
 of this universality class.

\section{Mapping to a gauge theory}
\begin{figure}
\includegraphics[width=0.9\hsize]{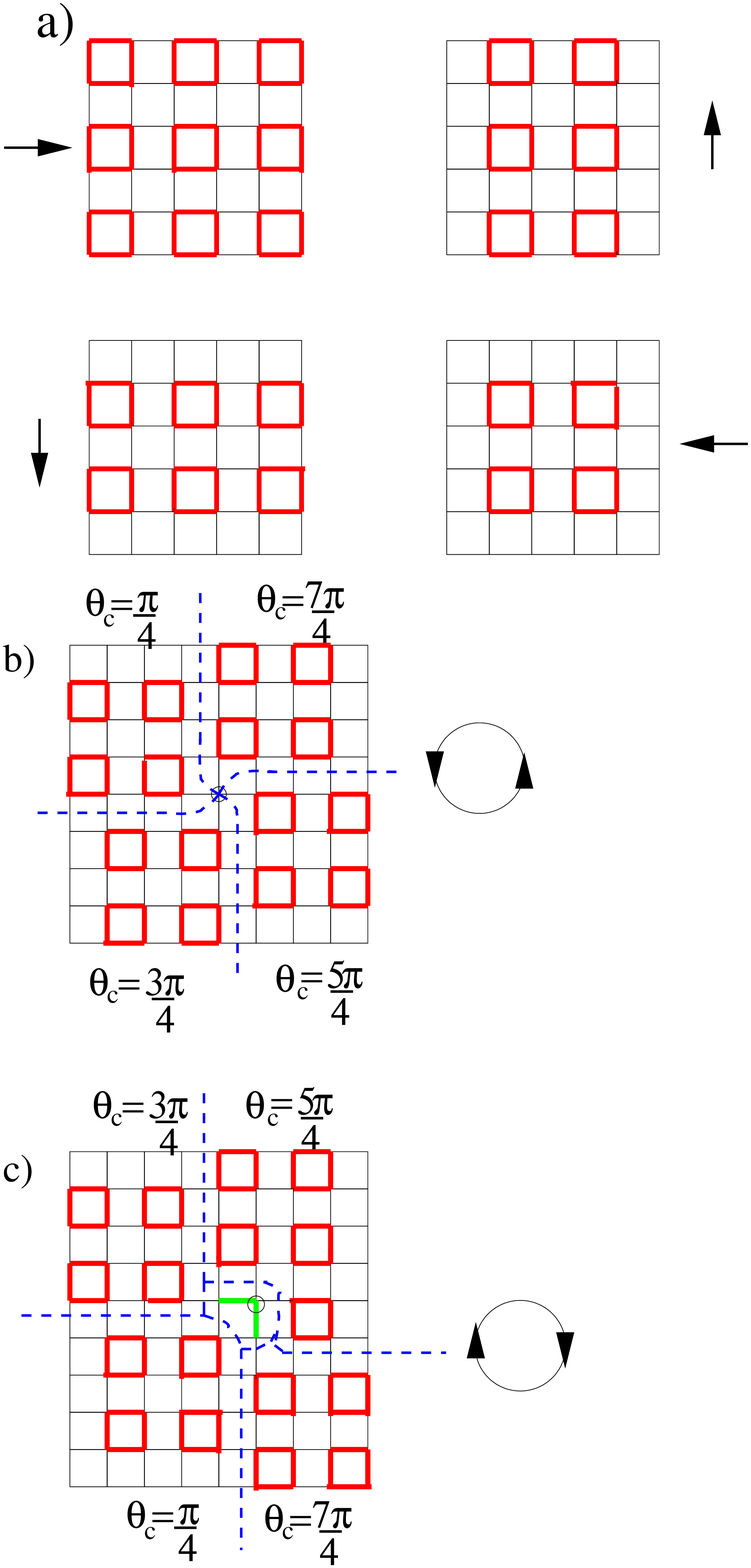}
      \caption{(Color online). a)The four degenerate ground states associated with the plaquette VBS state, which may be associated with four different orientations of a $Z_4$ order parameter $\psi_c$.Also shown in b) and c) are two types of $Z_4$ vortices in the plaquette state. The blue dotted lines represent the four domain walls. The encircled site is the core of the vortex.}
      \label{defect}
  \end{figure}
We now develop a theory for the phase transition of the model in terms of the defects of the Mott
insulator (MI) state at quarter filling, being careful to distinguish between transitions with
full low energy particle-hole symmetry, and those without. The insulator phase is characterized by a $Z_4$ clock order parameter $\psi_c$. In Fig~\ref{defect} (a), we show the four plaquette ground states which may be associated with the four orientations of the $Z_4$ order parameter (similarly, the four equivalent columnar states correspond to the four `columnar' choices of the phase of $\psi_c$). 

Because the order parameter is discrete, the natural topological defects are domain walls
between different orientations of the phase of $\psi_c$,  and four of them can terminate at a point to create a point-like $Z_4$ vortex excitation. There are two types of vortices present, depending on whether the core of the vortex is hole-like with a negative charge deviation at that
site (Fig~\ref{defect} (b)) or particle-like with a positive charge deviation (Fig~\ref{defect} (c)).
Translating the entire valence bond pattern by one lattice spacing reverses the direction of winding of the phase of the VBS order parameter. Thus, particle (hole)-like vortices have opposite windings on the two sublattices. Moreover, a particle-rich vortex has a phase winding which is
opposite to a hole-rich vortex located on the same sublattice.

Keeping these two points in mind, we define the charge of a vortex with its core
at site $r_i$ of the underlying square
lattice as $n_{r_i} \equiv (n_{1i}+n_{2i}+n_{3i}+n_{4i} - 1)$ for $r_i \in A$ sublattice,
and $n_{r_i} \equiv -(n_{1i}+n_{2i}+n_{3i}+n_{4i} - 1)$ for $r_i \in B$ sublattice. Note
that $n_{r_i}$ is non-zero only at the core of the vortices in any vortex configuration. Further,
since the total boson charge may be expressed as
\begin{equation}
Q= \frac{1}{2} \sum_i \left(n_{1i}+n_{2i}+n_{3i}+n_{4i} \right) \; ,
\end{equation}
we note that the vortices represent {\it fractionally charged excitations} carrying
quanta of charge equal
to one half the elementary boson charge. The mechanism for this is essentially
the same as in many
other examples discussed earlier in the literature ({\it e.g.} the Motrunich-Senthil
model realization of a $Z_2$ deconfined phase~\cite{Motrunich_Senthil}). However, in
contrast to earlier work on $Z_2$ deconfined phase, these fractionally charged vortex excitations are expected to be strongly interacting in our model (see below),  and we do {\it not} expect them to have an independent existence in the insulating phase. Nevertheless, as we will see below, they are convenient degrees of freedom in terms of which one may hope to describe the phase transition to a superfluid state, following the approach of Ref. \onlinecite{Levin_Senthil}.

In addition, we define $\tilde{n}_{r_i r_j}$ as the charge on the links of the square lattice where $r_i,r_j$ are nearest neighbours on the underlying square lattice, and identify it with the boson number at the center of the link. We also find it convenient to soften the hard-core constraint on the boson number, and instead capture the same physics through a strong on-site Hubbard potential. Finally, it is also useful to work with a rotor description of the charge degrees of freedom (measured from a background value corresponding to the quarter filled insulating state).
We can then describe the microscopic Hamiltonian in terms of rotor `angular momenta'  [$n_{r_A},n_{r_B},\tilde{n}_{r_A r_B}$] , and their conjugate angle variables [$\varphi_{r_A},\varphi_{r_B},\tilde{\varphi}_{r_A r_B}$].
Writing the effective Hamiltonian in terms of these variables, we obtain
\be
H &=& U\sum_{A}(n_{r_A}+1-f)^2 + U\sum_{B}(-n_{r_B}+1-f)^2 \nn \\
 &-&2t\sum_{r_A r_{A'}}\cos(\varphi_{r_A} - \varphi_{r_{A'}} + \tilde{\varphi}_{r_B r_A} - \tilde{\varphi}_{r_B r_{A'}})\nn \\
&-&2t\sum_{r_B r_{B'}}\cos(\varphi_{r_B} - \varphi_{r_{B'}} + \tilde{\varphi}_{r_A r_{B'}} - \tilde{\varphi}_{r_A r_B}) \nn \\
&+&V\sum_{r r'}\tilde{n}_{rr'}^2
\ee
along with the local constraints (due to the overcomplete set of variables used)
\be
\sum_{r_{B'}}\tilde{n}_{r_{A}r_{B'}} - 1 &=& n_{r_A} \nn \\
\sum_{r_{A'}}\tilde{n}_{r_{B}r_{A'}} - 1 &=& -n_{r_B}
\ee

As mentioned earlier, a large positive $V$ is introduced to make
the physics qualitatively similar to the hard-core boson case.
This Hamiltonian and the local constraints define a compact $U(1)$
gauge theory coupled to matter. To make this mapping explicit, we
define a compact vector potential $\vec{a}$ on the links of the
square lattice such that $\vec{a}_{r_A r_B} = \tilde{\varphi}_{r_A
r_B}$ and $\vec{a}_{r_A r_B} = -\vec{a}_{r_B r_A}$. The conjugate
electric field vector is then defined as $\vec{e}_{r_A r_B} =
\tilde{n}_{r_A r_B}$,  and $\vec{e}_{r_A r_B} = -\vec{e}_{r_B
r_A}$. The Hamiltonian in these new variables reads \be
\label{gaugerep}
H &=& U\sum_{A}(n_{r_A}+1-f)^2 + U\sum_{B}(-n_{r_B}+1-f)^2 \nn \\
&-&2t\sum_{r_A r_{A'}}\cos(\varphi_{r_{A'}} - \varphi_{r_A} - \int_{A}^{A'}\vec{a}\cdot d\vec{l})\nn \\
&-&2t\sum_{r_B r_{B'}}\cos(\varphi_{r_{B'}} - \varphi_{r_B} - \int_{B}^{B'}\vec{a}\cdot d\vec{l})\nn \\
&+&V\sum_{rr'}e_{rr'}^2 \; , \ee while the constraint can now be
written as $(\triangledown \cdot \vec{e})_r = n_r + \epsilon_r$,
where $\epsilon_A = 1$ and $\epsilon_B = -1$. Thus, our effective
model can be reformulated as a compact ($\vec{a}$ is an angle)
$U(1)$ lattice gauge theory coupled to two types of matter fields
$(n_{r_A},n_{r_B})$ in the presence of a background (static)
staggered charge pattern. 

The transformation properties of the
fields $\phi,n, a_{\alpha}, e_{\alpha}$ under discrete lattice
symmetries are readily obtained (see Table \ref{tab:psggt1}) and
play an important role in the discussion below. The global $U(1)$
symmetry associated with boson number conservation is realized by
the phase rotation
\begin{equation}
e^{i\phi} \rightarrow e^{i\alpha \epsilon_r} e^{i\phi}
\end{equation}
with $\alpha$ a constant. Correspondingly the total conserved
boson number is given by
\begin{equation}
Q = \frac{n_{r_A} - n_{r_B}}{2}
\end{equation}
Thus $e^{i\varphi_{r_A}}$ creates a state with total boson number
$1/2$ while $e^{i\varphi_{r_B}}$ creates a state with total boson
number $-1/2$.

\begin{table}[htbp]
  \centering
\vspace{0.5cm}
  \begin{ruledtabular}
  \begin{tabular}{l|c|c|c|c}
    Operation &  $\phi$  & $n$ & $a_{\alpha}$ & $e_{\alpha}$ \\
 \hline
 $T_x$ & $-\phi$ & $-n$ & $-a_{\alpha}$ & $-e_{\alpha}$ \\
 $R_{\pi/2}$ & $-\phi$ & $-n$ & $-\epsilon_{\alpha\beta} a_{\beta}$ & $-\epsilon_{\alpha\beta} e_{\beta}$ \\
$R_y$ & $ -\phi $ & $-n $ & $-a_{\alpha}$ & $-e_{\alpha}$\\
${\cal T}$ & $-\phi$ & $n$ & $-a_{\alpha}$ & $e_{\alpha}$
\end{tabular}
\end{ruledtabular}
\caption{Transformations of the fields on the square lattice under
the various independent discrete symmetry operations. $T_x$ is
translation by a unit cell in the $x$-direction. $R_{\pi/2}$ is
$\pi/2$ rotation about the center of a plaquette. $R_y$ is
reflection about the y-axis through midpoint of a horizontal bond.
${\cal T}$ is time reversal and is antiunitary. All other symmetry
operations may be built up from these. } \label{tab:psggt1}
\end{table}

This compact gauge theory is somewhat non-standard because the $A$
type particles can never convert to the $B$ type particles and
vice-versa, which makes it a bit inconvenient to analyse. To
overcome this technical difficulty, we define a theory with two kinds of matter
fields $(n_1,n_2)$ (and their conjugate variables
$(\varphi_1,\varphi_2)$), both of which can live on the $A$ and
$B$ sublattices, and hop between the $A$ and $B$ sublattice. The
original restriction that particles of one type only live on the corresponding
sublattice can be captured by a term of the kind
$(V\sum_{r_B}n_1^2 + V\sum_{r_A}n_2^2)$ with $V$ large and
positive as before. We expect that these changes preserve the universal
properties of our original theory.

Now, the charge at each site is $(n_{1r} - n_{2r})$, and the
constraint becomes $(\triangledown \cdot \vec{e})_r = n_{1r} +
n_{2r}+ \epsilon_r$. The electric field is however still related
to the microscopic boson density $n_{rr'}$ as $n_{rr'} =
\epsilon_r \vec{e}_{rr'}$ so that the electric field correlators
give direct information about the density correlations in our
microscopic model. Further simplification arises when we approach the
transition at the tip of the MI lobe. Because of particle-hole
symmetry at the tip, linear terms in $n_1,n_2$ vanish. As a result
it is enough for our purposes to consider the following simplified
theory: \be
H &=& U\sum_r (n_1^2 + n_2^2) - t'\sum_{r,\mu}\cos(\triangle_{\mu}\varphi_1 - a_{\mu}) \nn \\
&-&t'\sum_{r,\mu}\cos(\triangle_{\mu}\varphi_2 - a_{\mu}) + V\sum_{r,\mu}e_{\mu}^2 \nn \\
&-&K\sum_{\Box}\cos(\triangle \times \vec{a}) + V\sum_{r_B}n_1^2 + V\sum_{r_A}n_2^2 \; ; \nn \\
&& (\triangledown \cdot \vec{e})_r = n_{1r} + n_{2r}+ \epsilon_r
\ee where $\mu$ stands for the two lattice directions $\hat{x}$
and $\hat{y}$. The $K$ term has been added to make the theory look
like a more standard gauge theory and represents a ring-exchange
term in the dimer language. Such a term of this form would not
change any of the universal properties of the theory, and is expected to be dynamically generated
at low energies. The
transformation properties of these new fields under lattice
symmetries are also readily obtained and we list them below in
Table \ref{tab:psggt2}. As before global $U(1)$ symmetry is
realized by the phase rotation
\begin{eqnarray}
e^{i\phi_1} & \rightarrow & e^{i\alpha}e^{i\phi_1} \nn \\
e^{i\phi_2} & \rightarrow & e^{-i\alpha}e^{i\phi_2}
\end{eqnarray}

\begin{table}[htbp]
  \centering
\vspace{0.5cm}
  \begin{ruledtabular}
  \begin{tabular}{l|c|c|c|c}
    Operation &  $\phi_s$  & $n_s$ & $a_{\alpha}$ & $e_{\alpha}$ \\
 \hline
 $T_x$ & $-\phi_{\bar{s}}$ & $-n_{\bar{s}}$ & $-a_{\alpha}$ & $-e_{\alpha}$ \\
 $R_{\pi/2}$ & $-\phi_{\bar{s}}$ & $-n_{\bar{s}}$ & $-\epsilon_{\alpha\beta} a_{\beta}$ & $-\epsilon_{\alpha\beta} e_{\beta}$ \\
$R_y$ & $ -\phi_{\bar{s}} $ & $-n_{\bar{s}} $ & $-a_{\alpha}$ & $-e_{\alpha}$\\
${\cal T}$ & $-\phi_s$ & $n_s$ & $-a_{\alpha}$ & $e_{\alpha}$
\end{tabular}
\end{ruledtabular}
\caption{Transformations of the fields $(\phi_s,n_s, a_{\alpha},
e_{\alpha})$ on the square lattice under the various independent
discrete symmetry operations. Here $s = 1,2$ and $\bar{s} = 2$ if
$s = 1$ and vice-versa.} \label{tab:psggt2}
\end{table}

This model may now be analysed by a duality transformation. The results are rather similar to dual versions of boson models at 
half-filling on a square lattice. In particular there are two vortex species $\psi_1$ and $\psi_2$ (corresponding to vortices in $\phi_1$ and $\phi_2$ respectively). 
These two vortices move independently on the dual square lattice. The universal critical properties of the transition may then be described in a continuum 
description of this dual theory.
The oscillating background gauge charge (due to the $\epsilon_r$ in the Gauss law constraint) 
ensures that the leading term that allows for mixing of $\psi_1$ and $\psi_2$ vortices in the continuum dual action is
\begin{equation}
S_{mix} = -\int d^2x d\tau \lambda \left[\left (\psi_1^* \psi_2 \right)^4 + c.c \right]
\end{equation}
The full action then reads
\begin{eqnarray}
S & = & S_v + S_{mix} \nn \\
S_v & = & \int d^2x d\tau \sum_s |\left(\partial_{\mu} -i A_{\mu}\right)\psi_s|^2 + r|\psi_s|^2 \nn \\
& & +u_{d}|\psi_1|^2|\psi_2|^2 + u\left(|\psi_1|^2 + |\psi_2|^2 \right)^2 \nn \\
& & + \kappa \left(\epsilon_{\alpha\beta\gamma} \partial_\beta A_{\gamma} \right)^2
\end{eqnarray}
Here $A_{\mu}$ is the dual gauge field whose magnetic field strength is the original boson density. This is exactly the same critical theory as that derived in 
Ref. \onlinecite{courtney} to describe the superfluid-VBS transition of bosons at half-filling on a square lattice.

We thus expect that the present model lies in the same universality class as that transition. As argued in Ref. \onlinecite{Senthil_etal}, this is the easy plane NC{\it{CP}}$^1$ universality class and is described by the action $S_v$. The term $S_{mix}$ has been argued to be irrelevant at the corresponding fixed point. This low energy theory $S_v$  was
studied in Ref~\onlinecite{mv} by Motrunich and Vishwanath. Their analysis concluded that the theory was
self-dual and has a single continuous phase transition with unusual exponents that were
estimated in their work based on a detailed numerical study of a particular lattice regularization
of the theory. However, in more recent work, Kuklov {\em et. al.} have studied~\cite{Kuklov_etal} a different lattice regularization and
argued in favour of a weakly first order transition. Since the present work also finds a similar very weakly 
first order transition, more work appears to be needed to understand the physics of this universality class in greater
detail.

Finally, moving away from the tip of the Mott lobe corresponds to turning on a non-zero chemical potential in the original boson model. The transition is then expected to be either first order  or to proceed through two independent transitions with an intermediate supersolid phase. Although the present analysis cannot decide between these two possibilities without additional
input specific to the microscopic model being considered, our numerics provides evidence in favour of a first order
transition.

\section{Conclusions}
We have thus obtained convincing numerical evidence for a four-fold lattice symmetry breaking Mott insulator
at large boson repulsion in a simple model of hard-core bosons on the planar pyrochlore (checkerboard) lattice at quarter filling. We have also
studied the nature of the quantum phase transition from the superfluid state of the weakly interacting
system to the lattice symmetry broken Mott insulator. Our analytical results above suggest  a direct
transition at the tip of the $\rho = 1/4$ insulating lobe---more precisely, this transition would be
expected to be a continuous transition in the universality class of the easy-plane NC{\it{CP}}$^1$ model considered by Motrunich and Vishwanath in Ref~\onlinecite{mv}  

In our numerical work, we have studied the corresponding transition close to the tip of the lobe.
Our numerical evidence points to  a direct very weakly first order transition close to the tip of the lobe.
A detailed extrapolation of the strength of the first-order jump suggests that the transition remains
very weakly first order even at the particle-hole symmetric tip.
Thus, the present study underlines the need for further work to fully understand the physics
of the easy-plane NC{\it{CP}}$^1$ universality class, particularly with a view towards understanding
the very weakly first order nature of the transition seen in our work and in previous studies~\cite{Kuklov_etal}
of models that belong to this universality class.

\section{Acknowlegment}
We are grateful for access to computational resources at the Tata Institute of Fundamental Research and acknowledge support from the DAE Outstanding Investigator Program (TS) and DST SR/S2/RJN-25/2006 (KD).
.


\begin{thebibliography}{999}
\bibitem{Kanoda}
Y.~Kurosaki, Y.~Shimizu, K.~Miyagawa, K.~Kanoda and G.~Saito, Phys. Rev. Lett. {\bf 95}, 177001 (2005);
Y.~Shimizu, K.~Miyagawa, K.~Kanoda, M.~Maesato and G.~Saito, Phys. Rev. Lett. {\bf 91}, 107001 (2003).
\bibitem{Coldea}
R.~Coldea, D.~A.~Tennant and Z.~Tylczynski, Phys. Rev. B {\bf 68}, 134424 (2003).
\bibitem{Read_Sachdev} 
N.~Read and S.~Sachdev, Phys. Rev. Lett. {\bf 66}, 1773 (1991).
\bibitem{Wen} 
X.~G.~Wen, Phys. Rev. B {\bf 44}, 2664 (1991).
\bibitem{Senthil0}
T.~Senthil and M.~P.~A.~Fisher, Phys. Rev. B {\bf 62}, 7850 (2000).
\bibitem{Motrunich_Senthil}
O.~I.~Motrunich and T.~Senthil, Phys. Rev. Lett. {\bf 89}, 277004 (2002);
T.~Senthil and O.~I.~Motrunich, Phys. Rev. B {\bf 66}, 205104 (2002).
\bibitem{Balents_Fisher_Girvin}
L.~Balents, M.~P.~A.~Fisher and S.~M.~Girvin, Phys. Rev. B {\bf 65}, 224412 (2002).
\bibitem{Motrunich0}
O.~I.~Motrunich, Phys. Rev. B {\bf 67}, 115108 (2003).
\bibitem{Hermele_Balents_Fisher}
M.~Hermele, M.~P.~A.~Fisher and L.~Balents, Phys. Rev. B {\bf 69}, 064404 (2004).
\bibitem{Senthil_etal}
T.~Senthil {\it{et al.}}, Science {\bf 303}, 1490 (2004);
Phys. Rev. B {\bf 70}, 144407 (2004).
\bibitem{Levin_Senthil}
M.~Levin and T.~Senthil, Phys. Rev. B {\bf 70}, 220403 (2004).
\bibitem{Lee}
S.~S.~Lee and P.~A.~Lee, Phys. Rev. B {\bf 72}, 235104 (2005).
\bibitem{Melko0}
A.~W.~Sandvik, S.~Dual, R.~R.~P.~Singh and D.~J.~Scalapino, Phys. Rev. Lett. {\bf 89}, 247201 (2002).
\bibitem{Melko1}
R.~G.~Melko, A.~W.~Sandvik and D.~J.~Scalapino, Phys. Rev. B {\bf 69}, 100408 (R) (2004).
\bibitem{Melko2}
R.~G.~Melko and A.~W.~Sandvik, Phys. Rev. E {\bf 72}, 026702 (2005).
\bibitem{Sandviklatest} 
A.~W.~Sandvik, cond-mat/-0611343 (unpublished).
\bibitem{Damle_Senthil}
K.~Damle and T.~Senthil, Phys. Rev. Lett. {\bf 97}, 067202 (2006).
\bibitem{Isakov_Kim0}
K.~Sengupta, S.~V.~Isakov and Y.~B.~Kim, Phys. Rev. B {\bf 73}, 245103 (2006).
\bibitem{Isakov_Kim1}
S.~V.~Isakov, S.~Wessel, R.~G.~Melko, K.~Sengupta and Y.~B.~Kim, Phys. Rev. Lett. {\bf 97}, 147202 (2006).
\bibitem{Fulde1}
F.~Pollmann, P.~Fulde and E.~Runge, Phys. Rev. B {\bf 73}, 125121 (2006).
\bibitem{Fulde2}
F.~Pollmann and P.~Fulde, Europhys. Lett. {\bf 75}, 133 (2006).
\bibitem{Shtengel}
F.~Pollmann, J.~J.~Betouras, K.~Shtengel and P.~Fulde, Phys. Rev. Lett. {\bf 97}, 170407 (2006).
\bibitem{Nussinov1}
Z.~Nussinov, C.~D.~Batista, B.~Normand, and S.~A.~Trugman, cond-mat/0602528 (unpublished).
\bibitem{Nussinov2}
Z.~Nussinov, cond-mat/0606075 (unpublished).
\bibitem{Sachdev_Vojta}
S.~Sachdev and M.~Vojta, J. Phys. Soc. Japan, {\bf 69} Suppl. B, 1 (2000).
\bibitem{Syljuasen}
O.~F.~Syljuasen, Phys. Rev. B {\bf 73}, 245105 (2006).
\bibitem{Fisher_Fisher}
M.P.A. Fisher, P. Weichman, G. Grinstein and D.S. Fisher, Phys. Rev. B {\bf 40}, 546 (1989).
\bibitem{Supersolid1}
D.~Heidarian and K.~Damle, Phys. Rev. Lett. {\bf 95}, 127206 (2005).
\bibitem{Supersolid2}
R.~G.~Melko {\it{et al.}}, Phys. Rev.Lett. {\bf 95}, 127207 (2005).
\bibitem{Supersolid3}
S.~Wessel and M.~Troyer, Phys. Rev. Lett. {\bf 95}, 127205 (2005).
\bibitem{Supersolid4} 
M.~Boninsegni {\it{et al.}}, Phys. Rev. Lett. {\bf 97}, 080401 (2006).
\bibitem{Supersolid5}
P.~Sengupta {\it{et al.}}, Phys. Rev. Lett. {\bf 93}, 067003 (2004).
\bibitem{Supersolid6}
R.~G.~Melko, A.~Maestro and A.~A.~Burkov, cond-mat/060750 (unpublished).
\bibitem{SSEpre}
O.~Syljuasen and A.~Sandvik, Phys. Rev. E {\bf 66}, 046701 (2002).
\bibitem{SSEprb}
A.~Sandvik, Phys. Rev. B {\bf 59}, R14157 (1999).
\bibitem{SSEmath}
A.~Sandvik, J.Phys. A: Math. Gen. {\bf 25}, 3667 (1992).
\bibitem{courtney}
 C. Lannert, M. P. A. Fisher, and T. Senthil,
Phys. Rev. B 63, 134510 (2001) 
\bibitem{mv}
 O. I. Motrunich and A. Vishwanath,
Phys. Rev. B 70, 075104 (2004) 
\bibitem{Kuklov_etal} A.~Kuklov, N.~Prokof'ev, B.~Svistunov, and M.~Troyer,  cond-mat/0602466 (unpublished).

\end{thebibliography}
\end{document}